\RequirePackage{lineno}
\documentclass[aps,prl,twocolumn,superscriptaddress,floatfix,showpacs]{revtex4}
\usepackage{epsfig}
\usepackage{graphicx}
\usepackage{dcolumn}
\usepackage{amsmath}
\usepackage{xspace}
\usepackage{gensymb}
\graphicspath{{../PlotCollection/LongLiveAna/Results/}{../PlotCollection/EmTiming/EMTNIM/}
  {../PlotCollection/EmTiming/BeamHalo/}}

\begin{document}
\newcommand{\invpb}{$\rm pb^{-1}$} 
\newcommand{\invfb}{$\rm fb^{-1}$} 

\newcommand{\photon}{$\gamma$}
\newcommand{\ett}{\Et}
\newcommand{\ptt}{\mbox{$P_T$}}
\newcommand{\PT}{\ptt}
\newcommand{\NONE}{\mbox{$\widetilde{\chi}_1^0$}}
\newcommand{\NTWO}{\mbox{$\widetilde{\chi}_2^0$}}
\newcommand{\CONE}{\mbox{$\widetilde{\chi}_1^{\pm}$}}
\newcommand{\CONEP}{\mbox{$\widetilde{\chi}_1^{+}$}}
\newcommand{\CONEM}{\mbox{$\widetilde{\chi}_1^{-}$}}
\newcommand{\none}{\NONE}
\newcommand{\ntwo}{\NTWO}
\newcommand{\cone}{\CONE}
\newcommand{\coneb}{\CONEB}
\newcommand{\NTGNO}{\mbox{$\NTWO \rightarrow \gamma \NONE$}}
\newcommand{\NTGTGR}{\mbox{$\NONE \rightarrow \gamma \Gravitino$}}
\newcommand{\tanbeta}{\tan\beta}

\newcommand{\gevc}{GeV/c$^2$}

\newcommand{\mysect}[1]{\addtocounter{myctr}{1} \noindent \underline{Paragraph {\arabic{myctr}}: #1} \\}
\newcommand{\neut}{$ \tilde{\chi}_1^0 $}
\newcommand{\grav}{\ensuremath{\tilde{G}}}
\newcommand{\mett}{\mbox{$E\!\!\!\!/_{\rm T}$} }
\newcommand{\met}{\mett}
\newcommand{\eeggmett}{$ee\gamma\gamma\mett $}
\newcommand{\ggmett}{$\gamma\gamma\mett $}
\newcommand{\gmett}{$\gamma$\met}
\newcommand{\scinotn}[2]{\mbox{${#1}\times 10^{#2}$}}
\newcommand{\sumet}{$\sum E_{\rm T} $}
\newcommand{\sumpt}{$\sum P_{\rm T} $}
\newcommand{\dphicosm}{$\Delta\phi $}
\newcommand{\chisq}{${\chi}^2 $}
\newcommand{\etal}{{\it et al.}}

\newcommand{\zpos}{ {\it Z} }
\newcommand{\trkz}{$Z_{\rm 0}$}
\newcommand{\trkt}{$T_{\rm 0}$}

\def\roots{{\sqrt s}}
\def \Et {$E_{\rm T}$}
\def \Etg {E_{\rm T}^{\gamma}}
\def\Z{{ Z^0}}
\def\selectron{\widetilde{e}}
\def\goes{\rightarrow}
\def\Gravitino{\widetilde{G}}
\def\pbarp{p{\bar p}}
\def\degrees{^\circ}
\def\deg{^\circ}
\def\ppbar{p{\bar p}}

\def\W{{ $W^{\pm}$}}
\def\Wele{ $W \rightarrow e\nu$}
\def\Zee{ $Z \rightarrow e^+e^-$}

\def\Journal#1#2#3#4{{#1} {\bf #2} (#4) #3}
\def\PrePrint#1{\mbox{hep-ph/#1}}
\def\PRL{\rm Phys. Rev. Lett.}
\def\PRD{{\rm Phys. Rev.} D}
\def\NIM#1#2#3{\rm Nucl. Instr. and Meth A{#1}~(#2)~#3}

\def\rphi{r-$\phi$}                                                                           
\def\rz{r-z}
\def\and{\\ \vspace{0.3cm}}

\def\met{\mbox{$\not\!\!{E}_{T}\,$}}
\def\et{\mbox{$E_T$}}
\def\pt{\mbox{$p_T$}}
\def\tb{\mbox{$\tan\beta$}}
\def\mp{M_P}
\def\ms{{\bar M}_P}
\def\mdr{{\bar M}_D}
\def\md{M_D}
\def\Zee{\mbox{$Z\ra e e$}}
\def\Zmm{\mbox{$Z\ra \mu \mu$}}
\def\Ztt{\mbox{$Z\ra \tau \tau$}}
\def\Zll{\mbox{$Z\ra \ell \ell$}}
\def\Znn{\mbox{$Z\ra \nu \overline{\nu}$}}
\def\Wen{\mbox{$W\ra e\nu$}}
\def\Wmn{\mbox{$W\ra\mu\nu$}}
\def\Wtn{\mbox{$W\ra\tau\nu$}}
\def\Wln{\mbox{$W\ra\ell\nu$}}
\def\qqgG{\mbox{$q\ol{q} \ra gG$}}
\def\qgqG{\mbox{$qg \ra qG$}}
\def\gggG{\mbox{$gg \ra gG$}}

\def\ptmin{\mbox{$P_T^{\rm{min}}$}}

\title{\bf Search for large extra dimensions in final states containing one photon or jet and 
large missing transverse energy produced in $p\overline{p}$ collisions at $\sqrt{s} =$~1.96~TeV}
\preprint{FERMILAB-PUB-06/yyy-E}

\date{\today}

\begin{abstract}
We present the results of searches for large extra dimensions in samples of events 
with large missing transverse energy \mett and either a photon or a jet produced 
in $p\overline{p}$ collisions at $\sqrt{s}=1.96$~TeV collected with the CDF~II 
detector.  For $\gamma$ + \mett and jet + \mett candidate samples corresponding 
to 2.0~\invfb\ and 1.1~\invfb\ of integrated luminosity respectively, we observe
good agreement with standard model expectations and obtain a combined lower limit 
on the fundamental parameter of the large extra dimensions model, $M_{D}$, as a 
function of the number of extra dimensions in the model.   
\end{abstract}


\affiliation{Institute of Physics, Academia Sinica, Taipei, Taiwan 11529, Republic of China} 
\affiliation{Argonne National Laboratory, Argonne, Illinois 60439} 
\affiliation{University of Athens, 157 71 Athens, Greece} 
\affiliation{Institut de Fisica d'Altes Energies, Universitat Autonoma de Barcelona, E-08193, Bellaterra (Barcelona), Spain} 
\affiliation{Baylor University, Waco, Texas  76798} 
\affiliation{Istituto Nazionale di Fisica Nucleare Bologna, $^t$University of Bologna, I-40127 Bologna, Italy} 
\affiliation{Brandeis University, Waltham, Massachusetts 02254} 
\affiliation{University of California, Davis, Davis, California  95616} 
\affiliation{University of California, Los Angeles, Los Angeles, California  90024} 
\affiliation{University of California, San Diego, La Jolla, California  92093} 
\affiliation{University of California, Santa Barbara, Santa Barbara, California 93106} 
\affiliation{Instituto de Fisica de Cantabria, CSIC-University of Cantabria, 39005 Santander, Spain} 
\affiliation{Carnegie Mellon University, Pittsburgh, PA  15213} 
\affiliation{Enrico Fermi Institute, University of Chicago, Chicago, Illinois 60637} 
\affiliation{Comenius University, 842 48 Bratislava, Slovakia; Institute of Experimental Physics, 040 01 Kosice, Slovakia} 
\affiliation{Joint Institute for Nuclear Research, RU-141980 Dubna, Russia} 
\affiliation{Duke University, Durham, North Carolina  27708} 
\affiliation{Fermi National Accelerator Laboratory, Batavia, Illinois 60510} 
\affiliation{University of Florida, Gainesville, Florida  32611} 
\affiliation{Laboratori Nazionali di Frascati, Istituto Nazionale di Fisica Nucleare, I-00044 Frascati, Italy} 
\affiliation{University of Geneva, CH-1211 Geneva 4, Switzerland} 
\affiliation{Glasgow University, Glasgow G12 8QQ, United Kingdom} 
\affiliation{Harvard University, Cambridge, Massachusetts 02138} 
\affiliation{Division of High Energy Physics, Department of Physics, University of Helsinki and Helsinki Institute of Physics, FIN-00014, Helsinki, Finland} 
\affiliation{University of Illinois, Urbana, Illinois 61801} 
\affiliation{The Johns Hopkins University, Baltimore, Maryland 21218} 
\affiliation{Institut f\"{u}r Experimentelle Kernphysik, Universit\"{a}t Karlsruhe, 76128 Karlsruhe, Germany} 
\affiliation{Center for High Energy Physics: Kyungpook National University, Daegu 702-701, Korea; Seoul National University, Seoul 151-742, Korea; Sungkyunkwan University, Suwon 440-746, Korea; Korea Institute of Science and Technology Information, Daejeon, 305-806, Korea; Chonnam National University, Gwangju, 500-757, Korea} 
\affiliation{Ernest Orlando Lawrence Berkeley National Laboratory, Berkeley, California 94720} 
\affiliation{University of Liverpool, Liverpool L69 7ZE, United Kingdom} 
\affiliation{University College London, London WC1E 6BT, United Kingdom} 
\affiliation{Centro de Investigaciones Energeticas Medioambientales y Tecnologicas, E-28040 Madrid, Spain} 
\affiliation{Massachusetts Institute of Technology, Cambridge, Massachusetts  02139} 
\affiliation{Institute of Particle Physics: McGill University, Montr\'{e}al, Canada H3A~2T8; and University of Toronto, Toronto, Canada M5S~1A7} 
\affiliation{University of Michigan, Ann Arbor, Michigan 48109} 
\affiliation{Michigan State University, East Lansing, Michigan  48824}
\affiliation{Institution for Theoretical and Experimental Physics, ITEP, Moscow 117259, Russia} 
\affiliation{University of New Mexico, Albuquerque, New Mexico 87131} 
\affiliation{Northwestern University, Evanston, Illinois  60208} 
\affiliation{The Ohio State University, Columbus, Ohio  43210} 
\affiliation{Okayama University, Okayama 700-8530, Japan} 
\affiliation{Osaka City University, Osaka 588, Japan} 
\affiliation{University of Oxford, Oxford OX1 3RH, United Kingdom} 
\affiliation{Istituto Nazionale di Fisica Nucleare, Sezione di Padova-Trento, $^u$University of Padova, I-35131 Padova, Italy} 
\affiliation{LPNHE, Universite Pierre et Marie Curie/IN2P3-CNRS, UMR7585, Paris, F-75252 France} 
\affiliation{University of Pennsylvania, Philadelphia, Pennsylvania 19104} 
\affiliation{Istituto Nazionale di Fisica Nucleare Pisa, $^q$University of Pisa, $^r$University of Siena and $^s$Scuola Normale Superiore, I-56127 Pisa, Italy} 
\affiliation{University of Pittsburgh, Pittsburgh, Pennsylvania 15260} 
\affiliation{Purdue University, West Lafayette, Indiana 47907} 
\affiliation{University of Rochester, Rochester, New York 14627} 
\affiliation{The Rockefeller University, New York, New York 10021} 

\affiliation{Istituto Nazionale di Fisica Nucleare, Sezione di Roma 1, $^v$Sapienza Universit\`{a} di Roma, I-00185 Roma, Italy} 

\affiliation{Rutgers University, Piscataway, New Jersey 08855} 
\affiliation{Texas A\&M University, College Station, Texas 77843} 
\affiliation{Istituto Nazionale di Fisica Nucleare Trieste/\ Udine, $^w$University of Trieste/\ Udine, Italy} 
\affiliation{University of Tsukuba, Tsukuba, Ibaraki 305, Japan} 
\affiliation{Tufts University, Medford, Massachusetts 02155} 
\affiliation{Waseda University, Tokyo 169, Japan} 
\affiliation{Wayne State University, Detroit, Michigan  48201} 
\affiliation{University of Wisconsin, Madison, Wisconsin 53706} 
\affiliation{Yale University, New Haven, Connecticut 06520} 
\author{T.~Aaltonen}
\affiliation{Division of High Energy Physics, Department of Physics, University of Helsinki and Helsinki Institute of Physics, FIN-00014, Helsinki, Finland}
\author{J.~Adelman}
\affiliation{Enrico Fermi Institute, University of Chicago, Chicago, Illinois 60637}
\author{T.~Akimoto}
\affiliation{University of Tsukuba, Tsukuba, Ibaraki 305, Japan}
\author{M.G.~Albrow}
\affiliation{Fermi National Accelerator Laboratory, Batavia, Illinois 60510}
\author{B.~\'{A}lvarez~Gonz\'{a}lez}
\affiliation{Instituto de Fisica de Cantabria, CSIC-University of Cantabria, 39005 Santander, Spain}
\author{S.~Amerio$^u$}
\affiliation{Istituto Nazionale di Fisica Nucleare, Sezione di Padova-Trento, $^u$University of Padova, I-35131 Padova, Italy} 

\author{D.~Amidei}
\affiliation{University of Michigan, Ann Arbor, Michigan 48109}
\author{A.~Anastassov}
\affiliation{Northwestern University, Evanston, Illinois  60208}
\author{A.~Annovi}
\affiliation{Laboratori Nazionali di Frascati, Istituto Nazionale di Fisica Nucleare, I-00044 Frascati, Italy}
\author{J.~Antos}
\affiliation{Comenius University, 842 48 Bratislava, Slovakia; Institute of Experimental Physics, 040 01 Kosice, Slovakia}
% ...

\collaboration{CDF Collaboration\footnote{With visitors from $^a$Universiteit Antwerpen, B-2610 Antwerp, Belgium, 
$^b$Chinese Academy of Sciences, Beijing 100864, China, 
$^c$University of Bristol, Bristol BS8 1TL, United Kingdom, 
$^d$University of California Irvine, Irvine, CA  92697, 
$^e$University of California Santa Cruz, Santa Cruz, CA  95064, 
$^f$Cornell University, Ithaca, NY  14853, 
$^g$University of Cyprus, Nicosia CY-1678, Cyprus, 
$^h$University College Dublin, Dublin 4, Ireland, 
$^i$University of Edinburgh, Edinburgh EH9 3JZ, United Kingdom, 
$^j$Universidad Iberoamericana, Mexico D.F., Mexico, 
$^k$University of Manchester, Manchester M13 9PL, England, 
$^l$Nagasaki Institute of Applied Science, Nagasaki, Japan, 
$^m$University de Oviedo, E-33007 Oviedo, Spain, 
$^n$Queen Mary, University of London, London, E1 4NS, England, 
$^o$Texas Tech University, Lubbock, TX  79409, 
$^p$IFIC(CSIC-Universitat de Valencia), 46071 Valencia, Spain,
$^x$Royal Society of Edinburgh/Scottish Executive Support Research Fellow,
$^y$Istituto Nazionale di Fisica Nucleare, Sezione di Cagliari, 09042 Monserrato (Cagliari), Italy 
}}
\noaffiliation

\maketitle
\newcounter{myctr}



The mass of the Higgs particle in the standard model (SM) is subject to 
large quantum corrections. This is attributed to the existence of two 
equally fundamental energy scales in nature: the scale of the electroweak 
interaction ($\cal{O}$(100 GeV)) and the scale of the gravitational 
interaction ($\cal{O}$(10$^{19}$ GeV)).  One class of solutions to 
this hierarchy problem introduces new symmetries which protect physical 
parameters, such as the Higgs mass, from large quantum corrections. However, 
these models introduce an additional complication in that the new symmetries 
are required to be broken at some unknown scale and in some unknown way.  An 
alternate approach is to reconcile the hierarchy between the electroweak and 
gravity (Planck) scales by introducing extra spatial dimensions.


In the large extra dimensions (LED) scenario of Arkani-Hamed, Dimopoulos,
and Dvali~(ADD)~\cite{matchev}, gravity propagates in the $4+n$-dimensional 
``bulk'' of space-time, while the other SM fields are confined to our usual 
four dimensions.  The observed discrepancy between the size of Newton's 
constant and the strength of the electroweak couplings is understood as an 
artifact of the four-dimensional bias of the observer.  The four-dimensional 
Planck scale, $M_{Pl}$, is related to the fundamental $4+n$-dimensional 
Planck scale, $M_D$, by $M^2_{Pl} \sim R^n M_D^{n+2}$ where $n$ and $R$ are
the number and size of the extra dimensions respectively. 
An appropriate choice of $R$ for a given $n$ leads to a value of $M_D$
of the same order as that of the electroweak scale. 

Although models incorporating extra dimensions do not completely solve 
the hierarchy problem ($R$ has to be tuned to provide a match between 
the fundamental electroweak and Planck scales), their realization would 
provide an extraordinary and unique opportunity for direct studies of 
gravity at the Tevatron.  In these models the graviton ($G$ is used to 
denote all possible integer spin states from 0 to 2) is produced in the 
final states of the following interactions: $q\bar{q}\rightarrow\gamma 
G$, $q\bar{q}\rightarrow gG$, $qg\rightarrow qG$, and $gg\rightarrow gG$.  
The cross section for direct graviton production depends solely on the 
fundamental Planck scale $M_D$ due to cancellation of terms proportional
to $M_{Pl}$ (the relevant graviton-parton couplings suppress the cross 
section by $M_{Pl}^{-2}$ while the increased phase space volume due to   
the presence of the extra dimensions is proportional to $R^{n} \sim 
M^2_{Pl}/M_D^{n+2}$).
Conversely, the interaction of the produced graviton with material in 
the detector does not benefit from the increased phase space volume 
effect~\cite{grd}.  The final state graviton will therefore pass 
through the detector undetected, resulting in a signature of a single 
jet~\cite{jet} or photon accompanied by large missing transverse energy 
\mett~\cite{metdef}.  A previous search using single high-$E_T$ jet + 
\mett data corresponding to 368 pb$^{-1}$ of integrated luminosity 
collected with the Collider Detector at Fermilab (CDF~II)~\cite{monoPRL} 
observed no significant event excess with respect to SM expectations and 
placed the world's best lower limits on $M_D$ for the cases of five or 
more extra dimensions ($M_D >$~0.83~TeV for $n =$~6). 


In this Letter we describe an improved search for LED based on 
independent analyzes of single jet + \mett and $\gamma$ + \mett  
data samples corresponding to 1.1 and 2.0 fb$^{-1}$ of integrated
luminosity respectively.  Both analyzes focused on a specific event 
signature and were sensitive to a broad range of new physics models.  
Subsequent optimization of kinematic selection criteria was done 
to maximize sensitivity to the LED model.  The combined analysis 
presented here has significantly better sensitivity to LED model 
parameters than previous measurements.  Analysis of the $\gamma$ 
+ \mett data sample relies heavily on the electromagnetic shower 
timing system~\cite{nim}, which is critical in minimizing substantial 
cosmic ray backgrounds.

A full description of the CDF~II detector can be found elsewhere~\cite{CDFII}. 
Events in the $\gamma$ + \mett sample must satisfy the criteria at all three 
levels of the CDF~II trigger system for a high energy electromagnetic cluster 
($E_{T}>$~25~GeV) in the region $|\eta|<$~1.1 and \mett $>$~25~GeV.  The 
trigger is found to be $\sim$~100\% efficient for the final $\gamma$ and 
\mett kinematic selection requirements. The highest-\ett\ photon candidate 
in the fiducial region of the calorimeter is required to pass standard 
photon identification cuts~\cite{photons,ggmet}.  Candidate events are 
required to have $\mett >$ 50 GeV and contain at least one central photon 
with $|\eta|<$~1.1 and $E_T >$~50~GeV. To reduce $W$ + jet, where the 
jet is misidentified as a photon, and $W$ + $\gamma$ backgrounds events 
containing tracks with $p_T >$~10~GeV/c are vetoed.  We also reject events   
containing jets with $E_T >$ 15~GeV to reduce the background from $\gamma$
+ jet events with large \mett originating from jet energy mismeasurements.  
In order to reduce non-collision backgrounds we require a minimum of three 
good quality COT tracks in each candidate event.  The reconstructed photon 
is also required to be consistent in time with the $p\overline{p}$ collision 
and satisfy a discriminant~\cite{rvmmach} that separates photons produced in 
collisions from those originating from cosmic rays. 



Table~\ref{tab:backs} shows a breakdown of the estimated SM backgrounds 
in the $\gamma$ + $\mett$ event sample using two photon \ett\ 
requirements.  Collision-produced backgrounds include the irreducible 
contribution from $Z\gamma\rightarrow\nu\bar{\nu}\gamma$, $W\rightarrow 
\ell\nu$ production where the lepton is reconstructed as a photon, as 
well as $W\gamma$ and $\gamma\gamma$ production where the $W$ decay 
lepton or second photon is undetected.  The processes containing 
misidentified or undetected leptons are important at low energies 
but less so at higher energies since a small fraction of the leptons 
from $W$ decays are produced with $E_T >$~90~GeV.  The $Z\gamma$, 
$W\rightarrow\mu\nu$, and $W\rightarrow\tau\nu$ contributions are 
estimated from Monte Carlo simulation, while data-driven methods are 
used to estimate backgrounds for which the simulation is less reliable.

In the case of $W\rightarrow e\nu$ we rely on Monte Carlo simulation to 
determine the $E_T$ dependence of the probability for an electron to be 
reconstructed as a photon.  A data sample of $e\gamma$ events with small 
\mett and electron-photon invariant mass consistent with the $Z$ boson 
is used to normalize the modeled $E_T$ dependence.  A similar approach 
is used to estimate $W\gamma$ and $\gamma\gamma$ backgrounds.  The 
relative rate for observing only a single photon is determined from 
simulation, while the absolute normalization comes from the observed 
number of fully reconstructed events in data.  For example, we estimate 
the $\gamma\gamma$ background by determining the ratio of di-photon events 
with one and two reconstructed photons from simulation and multiplying by 
the number of observed two photon events in data.  Note that this approach 
also accounts for the additional contribution from $\gamma$ + jet events 
in cases where the original photon is lost and the jet is misidentified 
as a photon since the corresponding two photon events will be included 
in the event sample used for the normalization.

For photons with higher energies, the $Z\gamma\rightarrow\nu\nu\gamma$ 
background becomes increasingly dominant. To estimate this background we 
use a leading-order (LO) Monte Carlo simulation~\cite{baur}.  We determine 
that the LO description is adequate in the presence of our jet veto based 
on studies of the next-to-leading-order (NLO) version of the simulation, 
which indicate that the increase in the total cross section originating 
from the inclusion of NLO diagrams in the calculation is canceled by an 
equivalent decrease in acceptance due to the jet veto requirement.
  
\scriptsize
\begin{table}
\begin{center}
\begin{tabular}{ccc}
\hline\hline
Background & $E_{\rm T}^{\gamma} > 50$~GeV & $E_{\rm T}^{\gamma} > 90$~GeV \\ \hline
$W\rightarrow e\rightarrow\gamma$               & 47.3  $\pm$ 5.1  & 2.6  $\pm$ 0.4 \\
$W\rightarrow \mu/\tau\rightarrow\gamma$        & 19.1  $\pm$ 4.2  & 1.0  $\pm$ 0.2 \\
$W\gamma\rightarrow\mu\gamma\rightarrow\gamma$  & 33.1  $\pm$ 10.2 & 1.7  $\pm$ 1.2 \\
$W\gamma\rightarrow e\gamma\rightarrow\gamma$   & 8.0   $\pm$ 3.0  & 0.8  $\pm$ 0.7 \\
$W\gamma\rightarrow\tau\gamma\rightarrow\gamma$ & 17.6  $\pm$ 1.6  & 2.5  $\pm$ 0.2 \\
$\gamma\gamma\rightarrow\gamma$                 & 18.9  $\pm$ 2.3  & 2.3  $\pm$ 0.6 \\
Cosmic ray                                      & 36.4  $\pm$ 2.5  & 9.8  $\pm$ 1.3 \\
$Z\gamma\rightarrow \nu\nu\gamma$               & 100.1 $\pm$ 9.5  & 25.6 $\pm$ 2.0 \\ 
Total predicted                                 & 280.5 $\pm$ 15.7 & 46.3 $\pm$ 3.0 \\
Data observed                                   & 280              & 40             \\ 
\hline\hline
\end{tabular}
\end{center}
\caption{Number of observed events and expected SM backgrounds 
in the $\gamma$ + \mett candidate sample based on minimum photon 
$E_T$ requirements of 50 and 90~GeV.} 
\label{tab:backs}
\end{table}
\normalsize
 
Non-collision backgrounds which mimic the $\gamma$ + \mett signature 
originate from cosmic rays and particle interactions upstream of  
the detector.  Beam-produced muons traverse the calorimeter parallel 
to the beam line and deposit energy in multiple calorimeter towers 
covering the same azimuthal range.  Events with this topology are 
removed using cuts that minimally effect signal acceptance.

We use the calorimeter timing system to reduce background from 
cosmic rays.  Photon candidates originating from cosmic rays are 
uncorrelated in time with collisions and therefore produce roughly 
flat timing distributions.  The timing distribution of photons 
produced in collisions has a Gaussian shape with a mean of zero 
and standard deviation of 1.6~ns~\cite{nim}. We select photon 
candidates at least 20~ns out-of-time with the collision to predict 
the level of the background in the timing window around the collision.  
We also select a sample of pure cosmic photons to train a discriminant 
which separates collision photons from those produced by cosmic rays.  
The net effect of the discriminant is to reduce the cosmic ray 
background by a factor of $\sim$~600.  However, even with this large 
reduction factor, cosmic rays account for roughly 20\% of the total 
background in the high photon-$E_T$ region where we are most sensitive 
to new physics.

\begin{figure}[h]
  \centering%
  \includegraphics[width=0.8\linewidth]{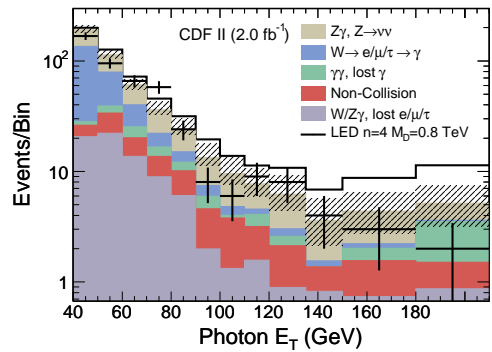}
  \caption[Predicted and observed photon $E_T$ distribution for $\gamma$
           + \mett events]
	  {Predicted (solid) and observed (crosses) photon $E_T$ 
           distributions for the $\gamma$ + \mett candidate sample.  
           The last bin shows all events containing a photon with 
           $E_T >$~180~GeV.  The expected LED signal for the case 
           of $n =$~4 and $M_D =$~0.8~TeV is also shown.  The 
           kinematic region above 90~GeV is used for constraining 
           the ADD model.  The hatched region indicates the total 
           uncertainty on the combined background prediction.} 
	  \label{fig:phoMet}
\end{figure} 

Figure~\ref{fig:phoMet} and Table~\ref{tab:backs} illustrate the 
agreement between the CDF~II $\gamma$ + \mett event sample and the 
SM background expectation.  We observe good agreement for both the 
low and high regions of the photon \Et\ spectrum.  

\begin{figure}[h]
  \centering%
  \includegraphics[width=0.8\linewidth]{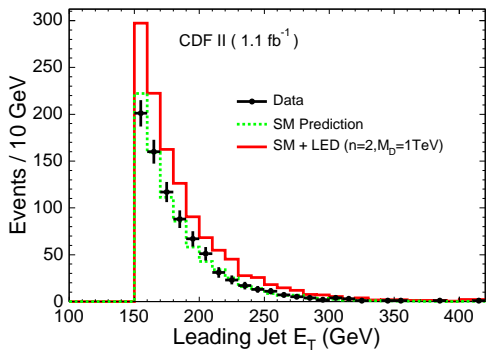}
  \caption{Predicted and observed leading jet $E_T$ distributions 
           for the jet + \mett candidate sample.  The expected 
           LED signal contribution for the case of $n =$~2 and 
           $M_D =$~1.0~TeV is also shown.}
	  \label{fig:jetet}
\end{figure}

The procedure used to analyze the jet + \mett sample has been described in a 
previous publication~\cite{monoPRL}.  The kinematic requirements, determined 
{\it a priori}, used to optimize sensitivity to LED are a single jet with 
$E_{T} >$~150~GeV and $\mett >$~120~GeV (a second jet with $E_T <$~60~GeV is 
allowed to increase signal acceptance).  The analysis reported here is simply 
an update to the previously published analysis.  The SM background estimates 
and the number of observed events are shown in Table~\ref{tab:jetback}, and 
a comparison of the expected and observed leading jet $E_T$ distributions is 
shown in Fig.~\ref{fig:jetet}.  

\scriptsize
\begin{table}
\begin{center}
\begin{tabular}{cc} 
\hline\hline
{Background}                      & {Events}       \\\hline
{$Z\rightarrow\nu\overline{\nu}$} & {388 $\pm$ 30} \\
{$W\rightarrow\tau\nu$}           & {187 $\pm$ 14} \\
{$W\rightarrow\mu\nu$}            & {117 $\pm$ 9}  \\
{$W\rightarrow e\nu$}             & {58 $\pm$ 4}   \\
{$Z\rightarrow \ell\ell$}         & {8 $\pm$ 1}    \\
{Multi-jet}                       & {23 $\pm$ 20}  \\
{$\gamma$+jet}                    & {17 $\pm$ 5}   \\
{Non-collision}                   & {10 $\pm$ 10}  \\
{Total predicted}                 & {808 $\pm$ 62} \\
{Data observed}                   & {809}          \\
\hline\hline
\end{tabular}
\end{center}
\caption{Number of observed events and expected SM backgrounds 
in the jet + \mett candidate sample.}
\label{tab:jetback}
\end{table}
\normalsize

Based on the observed agreement with the SM expectation in both the 
$\gamma$ + \mett and jet + \mett candidate samples, we proceed to 
set lower limits on $M_D$ for the LED model.  The limits are obtained 
solely from the total number of observed events in each of the samples 
(no kinematic shape information is incorporated). In order to estimate 
our sensitivity to the ADD model we simulate expected signals in both 
final states using the {\sc pythia}~\cite{pythia} event generator in 
conjunction with a {\sc geant}~\cite{geant} based detector simulation. 
For each extra dimension scenario we simulate event samples for $M_{D}$ 
ranging between 0.7 and 2~TeV.  In the case of the $\gamma$ + \mett 
analysis, the final kinematic selection requirements for the candidate 
sample are determined by optimizing the expected cross section limit 
without looking at the data.  The jet + \mett analysis was done as a 
generic search for new physics using three sets of kinematic cuts, the 
most sensitive of which is used here.  To compute the expected $95\%$ 
C.L. cross section upper limits we combine the predicted ADD signal and 
background estimates with systematic uncertainties on the acceptance 
using a Bayesian method with a flat prior~\cite{Conway}.  The acceptance 
is found to be almost independent (within 2\%) of the mass $M_{D}$.  
The total systematic uncertainties on the number of expected signal 
events are $5.7\%$ and $12.4\%$ for the $\gamma$ + \mett and jet + \mett 
candidate samples respectively.  The largest systematic uncertainties 
arise from modeling of initial/final state radiation convoluted with jet 
veto requirements, choice of renormalization and factorization scales, 
modeling of parton distribution functions, modeling of the jet energy 
scale (jet + \mett sample only), and the luminosity measurement.

\begin{figure}
  \centering%
  \includegraphics[width=0.8\linewidth]{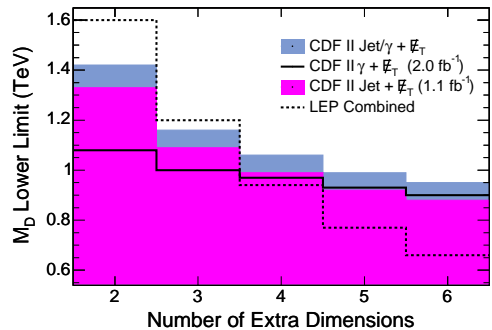}
  \caption{95~\% C.L. lower limits on $M_D$ in the ADD model as a function 
     of the number of extra dimensions in the model.}
  \label{fig:led}
\end{figure}

Since the underlying graviton production mechanism is equivalent for both 
final states, the combination of the independent limits obtained from the 
two candidate samples is based on the predicted relative contributions of 
the four graviton production processes.  Systematic uncertainties on the 
signal acceptances are treated as $100\%$ correlated, while uncertainties 
on background estimates, obtained in most cases from data, are considered 
to be uncorrelated.  The $95\%$ C.L. lower limits on $M_D$ from each 
candidate sample and the combined limits are given in Table~III and 
plotted with LEP limits~\cite{lep} in Fig.~\ref{fig:led}. 


\scriptsize
\begin{table}
\begin{center}
\begin{tabular}{cccccc}
\hline\hline
 & \multicolumn{2}{c}{$\gamma$ + \mett} & \multicolumn{2}{c}{Jet + \mett} & {Combined} \\
 $n$ &\ \  $\alpha$\ \ & $M_D^{obs}$ &\ \ $\alpha$\ \ & $M_D^{obs}$  & $M_D^{obs}$\\ \hline
 2 & 7.2 & 1080 & 9.9  & 1310 & 1400\\
 3 & 7.2 & 1000 & 11.1 & 1080 & 1150\\
 4 & 7.6 & 970  & 12.6 & 980  & 1040\\
 5 & 7.3 & 930  & 12.1 & 910  & 980\\
 6 & 7.2 & 900  & 12.3 & 880  & 940\\
\hline\hline
\end{tabular}
\label{tab:model}
\caption{Percentage of signal events passing the candidate sample selection 
criteria ($\alpha$) and observed $95\%$ C.L. lower limits on the effective 
Planck scale in the ADD model ($M_D^{obs}$) in GeV/c$^{2}$ as a function 
of the number of extra dimensions in the model ($n$) for both individual 
and the combined analysis.}
\end{center}
\end{table}
\normalsize

In conclusion, the CDF experiment has recently completed searches for new physics 
in the $\gamma$ + \mett and jet + \mett final states using data corresponding to 
2.0~\invfb\ and 1.1~\invfb\ of integrated luminosity respectively.  The observed 
number of events is consistent with the expected background in both channels, and 
we place limits on the ADD model of LED that are the world's best for the cases 
of four or more extra dimensions.  

\begin{acknowledgments}
We thank the Fermilab staff and the technical staffs of the
participating institutions for their vital contributions. This
work was supported by the U.S. Department of Energy and National
Science Foundation; the Italian Istituto Nazionale di Fisica Nucleare;
the Ministry of Education, Culture, Sports, Science and Technology 
of Japan; the Natural Sciences and Engineering Research Council of
Canada; the National Science Council of the Republic of China; the
Swiss National Science Foundation; the A.P. Sloan Foundation; the
Bundesministerium f\"ur Bildung und Forschung, Germany; the Korean
Science and Engineering Foundation and the Korean Research Foundation;
the Science and Technology Facilities Council and the Royal Society,
UK; the Institut National de Physique Nucleaire et Physique des 
Particules/CNRS; the Russian Foundation for Basic Research; the 
Ministerio de Educaci\'{o}n y Ciencia and Programa Consolider-Ingenio
2010, Spain; the Slovak R\&D Agency; and the Academy of Finland.
\end{acknowledgments}



\begin{thebibliography}{0}
\expandafter\ifx\csname natexlab\endcsname\relax\def\natexlab#1{#1}\fi
\expandafter\ifx\csname bibnamefont\endcsname\relax
  \def\bibnamefont#1{#1}\fi
\expandafter\ifx\csname bibfnamefont\endcsname\relax
  \def\bibfnamefont#1{#1}\fi
\expandafter\ifx\csname citenamefont\endcsname\relax
  \def\citenamefont#1{#1}\fi
\expandafter\ifx\csname url\endcsname\relax
  \def\url#1{\texttt{#1}}\fi
\expandafter\ifx\csname urlprefix\endcsname\relax\def\urlprefix{URL }\fi
\providecommand{\bibinfo}[2]{#2}
\providecommand{\eprint}[2][]{\url{#2}}

\end{thebibliography}


\begin{thebibliography}{99}



\bibitem{matchev} N.~Arkani-Hamed, S.~Dimopoulos,
  and G.~Dvali, Phys. Lett. B {\bf 429}, 263 (1998);
  Phys. Rev. D {\bf 59}, 086004 (1999).


\bibitem{grd} G.F. Giudice, R. Rattazzi, and J.D. Wells, 
  Nucl. Phys. B {\bf 544}, 3 (1999).

\bibitem{jet} Jets are reconstructed over the full pseudorapidity 
  coverage of the calorimeters ($|\eta| <$~3.6) using a fixed 
  cone algorithm and cone sizes $\Delta R = \sqrt{(\Delta\phi)^2 
  + (\Delta\eta)^2}$ of 0.4 and 0.7 for $\gamma$ + \mett and jet 
  + \mett events respectively.

\bibitem{metdef} We use a coordinate system where $\theta$ is the 
  polar angle to the proton beam, $\phi$ is the azimuthal angle 
  about the beam axis, and $\eta$ is the pseudorapidity defined as 
  $-{\rm ln~tan}(\theta/2)$.  Missing transverse energy \mett is 
  defined as $-\sum_{i} E_{T}^i \hat{n}^i$, where $\hat{n}^i$ is a 
  unit vector in the azimuthal plane from the beamline to the $i$th 
  calorimeter tower.

\bibitem{monoPRL} A. Abulencia {\it et al.} (CDF Collaboration),
  Phys. Rev. Lett. {\bf 97}, 171802 (2006).

\bibitem{nim} M.~Goncharov {\it et al.}, 
  Nucl. Instrum. Meth. A {\bf565}, 543 (2006). 

\bibitem{CDFII} D.~Acosta {\it et al.} (CDF Collaboration), 
  Phys. Rev. D \textbf{71}, 032001 (2005).












\bibitem{photons}
F. Abe {\it et al.} (CDF Collaboration), Phys. Rev. D {\bf 52}, 4784 (1995) 
contains electromagnetic cluster and photon identification variable definitions.


\bibitem{ggmet} 
F. Abe {\it et al.} (CDF Collaboration), Phys. Rev. Lett. {\bf 81}, 1791 (1998) 
contains standard photon identification criteria.  The $\chi^2_{CES}<20$ 
requirement is inefficient for photons incident at large angles relative to 
the detector face and therefore not used in selecting our $\gamma$ + \mett 
event sample.

\bibitem{rvmmach} We use a Relevance Vector Machine (RVM) trained on data to 
distinguish between collision and cosmic photon candidates (see C.C. Chang 
and C.J. Lin, ``{LIBSVM} : A Library for Support Vector Machines'' found at 
http://www.csie.ntu.edu.tw/$\sim$cjlin/libsvm).

\bibitem{baur}
U.~Baur, T.~Han, and J.~Ohnemus, Phys. Rev. D {\bf 57}, 2823 (1998).

\bibitem{pythia}
  T.~Sj\"{o}strand {\it et al.}, Comput. Phys. Commun. {\bf 135}, 238 (2001). 

\bibitem{geant} R.~Brun \etal, CERN-DD/EE/84-1 (1987).

\bibitem{Conway}
J.~Conway, CERN Yellow Book Report No. CERN 2000-005, 247 (2000).



\bibitem{lep} C. Ask, Proceedings of 32nd International Conference on High 
  Energy Physics (ICHEP 2004), Vol. {\bf 2}, 1289 (2005).


\end{thebibliography}
\end{document}